\documentclass[
  journal=largetwo,
  manuscript=article-type,
  year=2024,
  volume=3
]{cup-journal}

\newcount\ppnum
\newcommand\ppnumber[1]{%
        \ppnum=#1\relax
        \ifnum\ppnum<0
                $-$%
                \ppnum=-\ppnum
        \fi
        \let\pptemp\empty
        \loop\ifnum\ppnum>999
                \count255=\ppnum
                \divide\ppnum by1000
                \count255=\numexpr \count255 - 1000*\ppnum \relax
                \edef\pptemp{,\ifnum\count255<100 0\ifnum\count255<10 0\fi\fi
                             \the\count255 \pptemp}%
        \repeat
        \the\ppnum
        \pptemp
}

\usepackage[backend=biber]{biblatex}
\usepackage{amsmath, hhline}
\usepackage[nopatch]{microtype}
\usepackage{booktabs}
\usepackage{caption,subcaption}
\usepackage[]{hyperref}

\hypersetup{colorlinks = true, linkcolor = blue, anchorcolor = red, citecolor = blue, filecolor = red, urlcolor = red, hypertexnames=true}

\title{RS Sagittarii: Revealing the component spectra and the mass transfer}

\author{H. Bak{\i}\c{s}}
\affiliation{Department of Space Sciences and Technologies, Akdeniz University, 07058, Antalya, T\"{u}rkiye}
\email[H. Bak{\i}\c{s}]{hicranbakis@akdeniz.edu.tr}

\author{{\"{O}}.~Hilal Y{\i}ld{\i}z}
\affiliation{Graduate Institute of Natural and Applied Sciences, Akdeniz University, 07058, Antalya, T\"{u}rkiye}

\author{V. Bak{\i}\c{s}}
\affiliation{Department of Space Sciences and Technologies, Akdeniz University, 07058, Antalya, T\"{u}rkiye}

\author{G. Y\"{u}cel}
\affiliation{Department of Astronomy and Space Sciences, Istanbul University, 34119, Istanbul, T\"{u}rkiye}


\begin{document}

\begin{abstract}
We present an analysis of high-resolution ($R \sim 48000$) spectroscopic and photometric data of RS~Sgr, a short-period Algol-type binary system. For the first time, precise spectroscopic and absolute parameters of the system have been determined. The primary component is identified as a B3 main-sequence star with an effective temperature of 19000K, while the secondary is classified as an A0-type star with a temperature of 9700~K. The secondary appears to have recently evolved off the main sequence and currently fills its Roche lobe, transferring material through the inner Lagrangian point (L$_1$) to the hotter primary component.

The H$_\alpha$ emission and absorption features observed in the spectra are attributed to a combination of a low-density circumprimary disk, a gas stream originating from the secondary, and a hot spot formed at the impact site on the primary. The combined analysis of spectroscopic and photometric data yields a system distance of approximately 418~pc, which is consistent with the value derived from GAIA DR3 within the uncertainty limits.
\end{abstract}

\section{Introduction}

Considering that approximately half of the stars in the solar neighbourhood are binary or multiple systems, these systems are essential for understanding stellar evolution and testing theoretical stellar models. While considerable progress has been made for late-type binaries, the evolutionary pathways of interacting massive binary systems remain less well understood, primarily due to the limited number of such systems with well-constrained parameters. Accurate knowledge of the absolute parameters of the components is crucial for modelling mass transfer processes and accretion disk structures observed in these systems. Therefore, determining the absolute parameters of interacting massive binaries and multiple systems with high precision is a key goal in stellar astrophysics.

Such precision can be achieved by analysing high-quality light curves (LCs) obtained in multiple photometric bands, together with radial velocity (RV) curves, using modern modelling techniques. In this study, we analyse high-resolution spectroscopic data of RS~Sgr in combination with its multi-band LCs to determine accurate absolute parameters. The derived parameters are then compared with theoretical evolutionary models that take into account the binary nature of the system.

RS~Sgr (HD 167647, $V=6^{\mathrm{m}}.03$, $\alpha=18^{\mathrm{h}} 17^{\mathrm{m}} 36^{\mathrm{s}}$, $\delta=-34^{\circ}06' 26''$) is a southern Algol-type eclipsing binary that forms part of a multiple stellar system, accompanied by two visual companions: RS~Sgr-B ($V=9^{\mathrm{m}}.71$, $\alpha=18^{\mathrm{h}} 17^{\mathrm{m}} 40^{\mathrm{s}}$, $\delta=-34^{\circ} 06' 23''$) and RS~Sgr-C ($V=8^{\mathrm{m}}.76$, $\alpha=18^{\mathrm{h}} 17^{\mathrm{m}} 41^{\mathrm{s}}$, $\delta=-34^{\circ} 05' 15''$) \citep{Eggen_1982, Lindroos_1985}.

The photometric variability of RS~Sgr was first reported by \citet{Gould_1879}, and its Algol-type nature was later confirmed through time of minimum studies by \citet{roberts1895, roberts1896, roberts1901}. \citet{1948MNRAS.108..343B}, using the LCs obtained by \citet{redman1945}, modeled the system’s orbit as circular. The orbital period was refined to 2.41568311 days by \citet{1990AcA....40..283C}, who also derived the absolute parameters of the components. Their analysis indicated that the primary is a main-sequence star, while the secondary is less massive, cooler, nearly filling its Roche lobe, and consequently over-luminous.

In contrast, \citet{1980ApJ...237..513M} proposed an eccentric orbit for the system and reported an apsidal motion period of approximately 27 years. The system is also listed with an eccentricity of $e = 0.008$ in the catalog of eccentric binaries by \citet{1999AJ....117..587P}.

Several photometric studies of RS~Sgr exist in the literature \citep[e.g.,][]{1969ApJ...157..313H, 1971AJ.....76..621C, 1983ApJS...51..321S, 1983ApJS...52..429W, 1987AJ.....94..771D}, though complete LCs are only available from \citet{shobbrook2004}, the {\it Hipparcos} ({\it HIP}) mission \citep{ESA}, and the {\it TESS} space telescope \citep{ricker2015}.

In the present study, high-resolution spectroscopic data of \citet{bakics2010spectroscopic} are re-investigated together with available photometric data in the literature to determine the nature of the interacting binary system RS Sgr.

\section{Observations}
\label{sec:observations}
\subsection{Photometric Data}

Photometric observations of the RS~Sgr system have been obtained from both ground-based facilities and space missions. The {\it Hipparcos} satellite \citep{ESA} monitored RS Sgr between 1990 and 1993, collecting a total of 79 data points with a mean standard error of 0.006 mag. However, it is noteworthy that the number of measurements obtained near the primary minimum is limited.

The system was also observed by the {\it Transiting Exoplanet Survey Satellite (TESS)}\footnote{\url{https://www.nasa.gov/tess-transiting-exoplanet-survey-satellite}}. For the present analysis, we used the {\it SPOC} pipeline data from sector 66 with a 120-second cadence. The LC was normalised to unity prior to the analysis. The observation data were retrieved from the Mikulski Archive for Space Telescopes (MAST)\footnote{\url{https://mast.stsci.edu/}}.

\begin{figure}
 \resizebox{80mm}
   {!}{\includegraphics{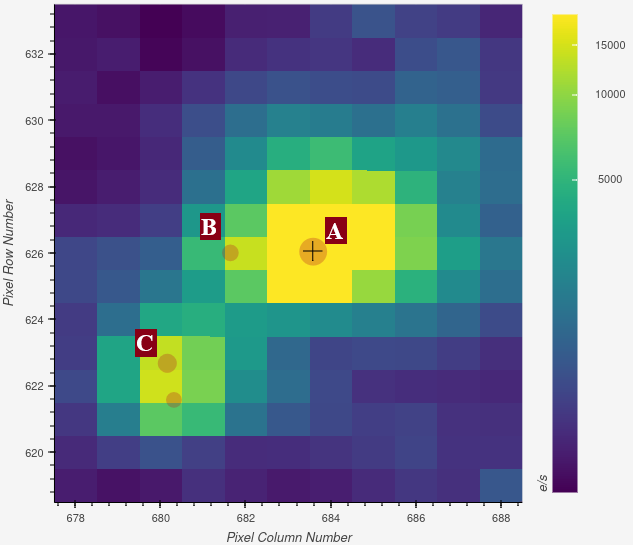}}
    \caption{The TESS band image of RS Sgr and other stars in its vicinity. The image dimension is 3.85x5.25 arcmin$^{2}$.}
    \label{fig:tessimage}
\end{figure}

Figure~\ref{fig:tessimage} displays the TESS-band image of the system in a $3.85 \times 5.25$ arcmin$^{2}$ field centered on RS~Sgr (labelled A). The nearest visual companion, RS~Sgr-B (labelled B), is located at a projected angular separation of approximately 40 arcseconds, while RS~Sgr-C (labelled C) lies at roughly 118 arcseconds. The brightness limit of the field is around 12 mag in the G-band.

The aperture size used in the TESS SPOC pipeline is determined based on the target's brightness. For RS~Sgr, this corresponds to an aperture radius of 3 pixels, or about 31 arcseconds (since one TESS pixel spans 21 arcseconds). Given both the angular separation ($\sim$40 arcseconds) and the relative faintness ($V = 9.7$ mag) of RS~Sgr-B, its light contribution is assumed to be negligible in the extracted LC.

Ground-based photometric data were obtained by \citet{shobbrook2004} using a 24-inch (61 cm) telescope of the Australian National University at Siding Spring Observatory (SSO) between the years 1991-2001. A total of 110 measurements were performed using the Johnson-V filter. Photometric errors in measurements were typically 0.001 to 0.003 mag.

\subsection{Spectroscopic Data}

High-resolution spectroscopic observations of RS~Sgr were carried out using the 1-m McLennan Telescope equipped with the High Efficiency and Resolution Canterbury University Large Echelle Spectrograph (HERCULES) at Mt. John University Observatory, New Zealand \citep{Hearnshaw2002, Hearnshaw-etal2002}. HERCULES is a fibre-fed echelle spectrograph housed in a thermally insulated environment, providing stable observational conditions. It offers broad wavelength coverage from 3800 to 8800\AA \ across more than 80 echelle orders. The instrument supports two resolving powers, $R = 41000$ and $R = 70000$, selectable via three interchangeable optical fibres of varying diameters.

A total of 35 high-resolution spectra of RS~Sgr were obtained by \citet{bakics2010spectroscopic}, and these constitute the spectroscopic dataset used in this study. All observations were carried out with the 4k$\times$4k Spectral Instruments 600S CCD camera. The data were reduced using the HERCULES Reduction Software Package (HRSP; \citealp{skuljan2004hrsp}) following standard echelle spectroscopic reduction procedures. Details of the observations, including the observing dates, exposure times, mean signal-to-noise ratios at 5500\AA, and orbital phases, are listed in Table~\ref{tab:information}.

\begin{table*}
	\centering
	\caption{The log of spectroscopic observations together with measured RVs and their standard errors. The RVs given are relative to $V_\gamma$. The S/N ratio refers to 5500 \AA.}
	\label{tab:information}
	\begin{tabular}{llccccccc}
		\hline 
		\hline
		No  & HJD & Exp. Time & S/N & Phase &  RV$_1$ & errRV$_1$ & RV$_2$ & errRV$_2$ \\
	& (- 2\,400\,000) & (s) & & & (km s$^{-1}$) & (km s$^{-1}$) & (km s$^{-1}$) & (km s$^{-1}$) \\
		\hline 
		\hline 
		\hline 
$1$&$	53970.8574	$&$	289	$&$	95	$&$	0.88	$&$	58.87	$&$	0.00	$&$	-181.35	$&$	0.00	$\\
$2$&$	53972.0517	$&$	699	$&$	88	$&$	0.37	$&$	-61.91	$&$	-0.75	$&$	187.78	$&$	-0.64	$\\
$3       $&$	53972.0933	$&$	642	$&$	78	$&$	0.39	$&$	-52.59	$&$	1.35	$&$	165.71	$&$	-0.46	$\\
$4$&$	53978.0006	$&$	543	$&$	78	$&$	0.84	$&$	74.56	$&$	0.20	$&$	-228.45	$&$	0.62	$\\
$5$&$	53978.0573	$&$	893	$&$	78	$&$	0.86	$&$	66.71	$&$	0.14	$&$	-204.88	$&$	0.19	$\\
$6$&$	53981.0393	$&$	766	$&$	84	$&$	0.09	$&$	-51.36	$&$	-0.22	$&$	157.32	$&$	-0.22	$\\
$7$&$	53983.8032	$&$	422	$&$	92	$&$	0.24	$&$	-87.91	$&$	0.26	$&$	271.57	$&$	-0.03	$\\
$8	$&$	53983.8086	$&$	402	$&$	89	$&$	0.24	$&$	-87.90	$&$	0.32	$&$	271.58	$&$	-0.19	$\\
$9	$&$	53983.8373	$&$	485	$&$	86	$&$	0.25	$&$	-87.90	$&$	0.29	$&$	271.56	$&$	-0.13	$\\
$10$&$	53983.8439	$&$	524	$&$	85	$&$	0.26	$&$	-87.89	$&$	0.22	$&$	271.57	$&$	0.11	$\\
$11$&$	53983.8817	$&$	628	$&$	86	$&$	0.27	$&$	-87.89	$&$	-0.70	$&$	269.01	$&$	0.42	$\\
$12$&$	53983.9239	$&$	1009	$&$	74	$&$	0.29	$&$	-85.00	$&$	0.16	$&$	261.38	$&$	-0.97	$\\
$13$&$	53983.9359	$&$	948	$&$	75	$&$	0.29	$&$	-84.78	$&$	-0.38	$&$	261.25	$&$	1.26	$\\
$14$&$	53984.8090	$&$	1204	$&$	112	$&$	0.65	$&$	73.01	$&$	-1.14	$&$	-228.42	$&$	-0.01	$\\
$15$&$	53984.8610	$&$	843	$&$	120	$&$	0.68	$&$	79.28	$&$	-0.65	$&$	-246.15	$&$	0.07	$\\
$16$&$	53984.8936	$&$	1076	$&$	113	$&$	0.69	$&$	82.42	$&$	-0.39	$&$	-254.95	$&$	0.16	$\\
$17$&$	53984.9284	$&$	689	$&$	105	$&$	0.70	$&$	85.55	$&$	0.32	$&$	-262.95	$&$	-0.38	$\\
$18$&$	53984.9787	$&$	713	$&$	112	$&$	0.73	$&$	87.21	$&$	-0.29	$&$	-269.27	$&$	0.28	$\\
$19$&$	53984.9875	$&$	653	$&$	104	$&$	0.73	$&$	87.13	$&$	-0.62	$&$	-270.79	$&$	-0.49	$\\
$20$&$	53985.0238	$&$	764	$&$	106	$&$	0.74	$&$	87.13	$&$	-1.13	$&$	-270.87	$&$	1.02	$\\
$21$&$	53985.0337	$&$	813	$&$	102	$&$	0.75	$&$	88.66	$&$	0.39	$&$	-270.79	$&$	1.11	$\\
$22$&$	53985.0701	$&$	716	$&$	95	$&$	0.76	$&$	88.69	$&$	0.91	$&$	-269.26	$&$	1.14	$\\
$23$&$	53985.0793	$&$	744	$&$	91	$&$	0.77	$&$	87.13	$&$	-0.40	$&$	-269.25	$&$	0.39	$\\
$24$&$	53985.9348	$&$	648	$&$	118	$&$	0.12	$&$	-63.57	$&$	-1.27	$&$	190.74	$&$	-1.20	$\\
$25$&$	53986.0104	$&$	665	$&$	109	$&$	0.15	$&$	-74.54	$&$	-1.23	$&$	225.27	$&$	-0.60	$\\
$26$&$	53986.0425	$&$	723	$&$	96	$&$	0.17	$&$	-77.71	$&$	-0.55	$&$	236.29	$&$	-1.42	$\\
$27$&$	53988.8602	$&$	508	$&$	109	$&$	0.33	$&$	-74.58	$&$	1.22	$&$	233.13	$&$	-0.38	$\\
$28$&$	53989.0159	$&$	568	$&$	98	$&$	0.40	$&$	-51.07	$&$	0.78	$&$	159.46	$&$	-0.28	$\\
$29$&$	53989.9563	$&$	516	$&$	101	$&$	0.79	$&$	87.13	$&$	1.54	$&$	-262.96	$&$	0.69	$\\
$30$&$	53989.9927	$&$	540	$&$	99	$&$	0.80	$&$	83.99	$&$	0.82	$&$	-255.11	$&$	1.08	$\\
$31$&$	53990.0270	$&$	554	$&$	99	$&$	0.81	$&$	80.82	$&$	0.61	$&$	-245.71	$&$	1.36	$\\
$32$&$	53990.9697	$&$	1480	$&$	87	$&$	0.21	$&$	-85.55	$&$	-0.23	$&$	262.89	$&$	0.04	$\\
$33$&$	53991.8993	$&$	499	$&$	105	$&$	0.59	$&$	48.76	$&$	-0.20	$&$	-151.48	$&$	-0.68	$\\
		\hline 
		\hline 
		\end{tabular}
\end{table*}

\section{Spectroscopic Orbit}

RV measurements of RS~Sgr were performed on normalised spectra using the spectral disentangling (SD) technique. The orbital solution parameters derived from the cross-correlation method by \citet{bakics2010spectroscopic} were adopted as initial inputs for the SD analysis. These parameters were then refined through the SD technique to obtain the final spectroscopic orbital solution.

The orders 97, 113, and 127, which were previously analysed using the cross-correlation (CC) method \citep{bakics2010spectroscopic}, were also examined using the SD method with the updated version of the \textsc{\large korel} code\footnote{https://stelweb.asu.caz.cz/vo-korel/} \citep{hadrava2004}. This code provides the orbital parameters and disentangled spectra of the component stars through the Fourier disentangling method. The \textsc{\large korel} tool is particularly valuable for binary stars exhibiting shallow and/or blended spectral lines, such as R~Ara \citep{2016MNRAS.458..508B} and HH~Car \citep{2021MNRAS.503.2432B}. Before initiating the solutions with \textsc{\large korel}, it is essential to analyse the LCs of the system and determine the phase-dependent light contributions of the components for an accurate reconstruction of the component spectra. To achieve this, the light curves of RS~Sgr were carefully analysed to establish the light contributions of the components at the relevant wavelengths and phases, as described in Section 4.

In the solutions of the RS~Sgr system, the orbital period is taken as 2$^d$.4156835 \citep{kreiner2004} and it is fixed. Other parameters such as ephemeris time, $T_{0}$, eccentricity, $e$, periastron longitude, $\omega$, RV semi-amplitude of the primary component, $K_{1}$ and mass ratio, $q$ are set as free parameters during the orbit determination. The centre of mass velocity cannot be determined with the code-\textsc{\large korel}, therefore, the RVs measured with this code are only relative to the $V_\gamma$. The $V_\gamma$ velocity of the RVs in Table \ref{tab:information} and Fig. \ref{fig:rvcurve} is zero. However, in our previous study \citep{bakics2010spectroscopic}, the $V_\gamma$ velocity was calculated as --4.4 $\pm$ 1.0 km/s, which should be added to the RVs in Table \ref{tab:information} for their absolute values. The uncertainties in the measured RVs are adopted as the standard deviations (STD) of the RVs obtained from three spectral orders (97, 113, and 127). The RVs with their standard errors (SE), calculated as $SE=STD/\sqrt{N}$ with N=3, for each component, along with the final orbital solution parameters of the system derived using the code \textsc{\large KOREL}, are presented in Table~\ref{tab:information} and Table~\ref{tab:rvsolution}, respectively.

 \begin{table}
 	\caption{The final parameters for the spectroscopic orbit of the  RS~Sgr.}	
 	\begin{tabular}{lc}
 		\hline 
 		\hline
    \textbf{Parameter(Unit)} & \textbf{Value}  \\ 
 		\hline 
 		\hline 
 		$P$ (d) & 2.4156835  \\
 		$T_0$ (d) & 2452503.587 $\pm$0.003 \\
 		$K_1$ (km/s)& 88.34 $\pm$0.13  \\
 		$K_2$ (km/s) & 271.53 $\pm$0.13  \\
 		$q$  & 0.325 $\pm$0.001\\
 		$e$  & 0.0 \\
 		$a_1sini$ (\(\textup{R}_\odot\)) & 4.192 $\pm$0.006  \\
 		$a_2sini$ (\(\textup{R}_\odot\)) & 12.885 $\pm$0.006  \\
 		$m_1sin^3i$ (\(\textup{M}_\odot\)) & 8.801 $\pm$0.004 \\
 		$m_2sin^3i$ (\(\textup{M}_\odot\)) & 2.863 $\pm$0.004  \\
 		\hline 
 		\hline 
 		\label{tab:rvsolution}
 	\end{tabular}
 
 \end{table}

\section{Analysis of the Photometric Data}

The photometric data compiled from \citet{shobbrook2004}, as well as from the {\it HIP} and {\it TESS} missions, were analysed using the Wilson-Devinney code \textsc{\large (wd)} \citep{1971ApJ...166..605W, 1979ApJ...234.1054W, 1990ApJ...356..613W, van2007third, 2008ApJ...672..575W, 2010ApJ...723.1469W, wilson2012spotted, wilson2013unification} through the {\sc phoebe} graphical user interface\footnote{https://github.com/phoebe-project/phoebe1.git} for LC modelling. RS~Sgr is a semi-detached eclipsing binary system, with the secondary component filling its Roche lobe and transferring mass to the primary. Accordingly, the analysis was conducted under this configuration. The orbital period ($P$) and the reference epoch ($T_{0}$) were adopted from \citet{kreiner2004}, where $P$ was fixed and $T_{0}$ was allowed to converge during the fitting process. The semi-major axis, mass ratio (from the spectroscopic solution), and the surface potential of the secondary component were held constant throughout the modelling. The initial value of the orbital inclination ($i = 82.5^\circ$) was adopted from \citet{1990AcA....40..283C} and treated as an adjustable parameter.

One of the most critical fixed parameters in LC analysis is the effective temperature of one of the stellar components. For early-type stars, the effective temperature can be reliably estimated using photometric colours via the $Q$-method introduced by \citet{johnson1953fundamental}. The standard colours and $V$-band brightness of RS~Sgr, reported by \citet{1983ApJS...51..321S}, are $U-B = -0.63$, $B-V = -0.08$, and $V = 6.01$. An average value of $E(U-B)/E(B-V) = 0.72$ was adopted in the $Q$-method calculation. Accordingly, the derived $Q$ parameter for the system is $-0.572$.

Using this $Q$ value, the intrinsic (unreddened) colour indices were calculated as $(B-V)_0 = -0.202$ and $(U-B)_0 = -0.718$. Based on these values, the effective temperature of the primary component was estimated to be approximately 20,000 K using the calibration tables presented in \citet{Eker2020}. This temperature was adopted as a fixed input for the primary star, while the temperature of the secondary component was treated as a free parameter and allowed to converge during the LC modelling.

At each iteration of the solution process, logarithmic limb-darkening coefficients for the bolometric, {\it HIP} (490 nm), {\it TESS} (786.5 nm), and $V$ (550 nm) bands were adopted from the tables of \citet{van1993new}. The bolometric albedos and gravity-darkening exponents for both components were taken as $A_{1,2} = 1$ and $g_{1,2} = 1$, consistent with radiative envelope conditions \citep{Rafert1980}.

As a result of the initial analysis based on the colour-derived temperature, the best-fitting LC models were obtained. Using this model, the phase-dependent light contributions of the individual components were determined for the 4900~\AA\ ({\it HIP} band), 5500~\AA\ ($V$ band), and 7865~\AA\ ({\it TESS} band) photometric passbands. These light contributions were subsequently adopted as input parameters in the \textsc{\large korel} code to reconstruct the disentangled spectra of the stellar components within the wavelength intervals described in Section~5.

The effective temperatures of the primary and secondary components were also determined through atmospheric modelling of the disentangled spectra, yielding adopted values of $T_\mathrm{eff,1} = 19000$~K and $T_\mathrm{eff,2} = 9500$K, respectively (see Section 5). In the subsequent photometric analysis, the primary's temperature and the spectroscopically derived mass ratio were fixed parameters. The analysis was repeated following the same approach.

From the atmosphere models, the projected rotational velocities ($v \sin i$) of the primary and secondary components were derived as 115 kms$^{-1}$ and 90 kms$^{-1}$, respectively. These values are consistent, within uncertainties, with the expected synchronous rotation rates. Therefore, the synchronicity parameter $F$ (the ratio of observed to synchronous rotational velocity) was set to 1 for both components during the LC analysis.

The {\it HIP} and Shobbrook (2004) $V$-band LCs were analysed simultaneously, while the {\it TESS} LC was modelled separately under the same initial conditions. From these refined solutions, the temperature of the secondary component was estimated as approximately 9700K. The final photometric parameters, derived from the combined LC and RV analysis, are summarised in Table \ref{tab:LC}.

To refine the component parameters and estimate their uncertainties, a Markov Chain Monte Carlo (MCMC) sampler based on the \texttt{emcee} framework \citep{2013PASP..125..306F} was employed\footnote{\url{https://sourceforge.net/p/phoebe/mailman/message/33650955/}}. The MCMC computation was performed using 128 walkers and 1000 iterations. Due to limited phase coverage in the {\it HIP} and Shobbrook (2004) data, the MCMC method was only applied to the {\it TESS} dataset.

\begin{figure}
\begin{subfigure}[htbp]{0.8\linewidth}
\includegraphics[width=\linewidth]{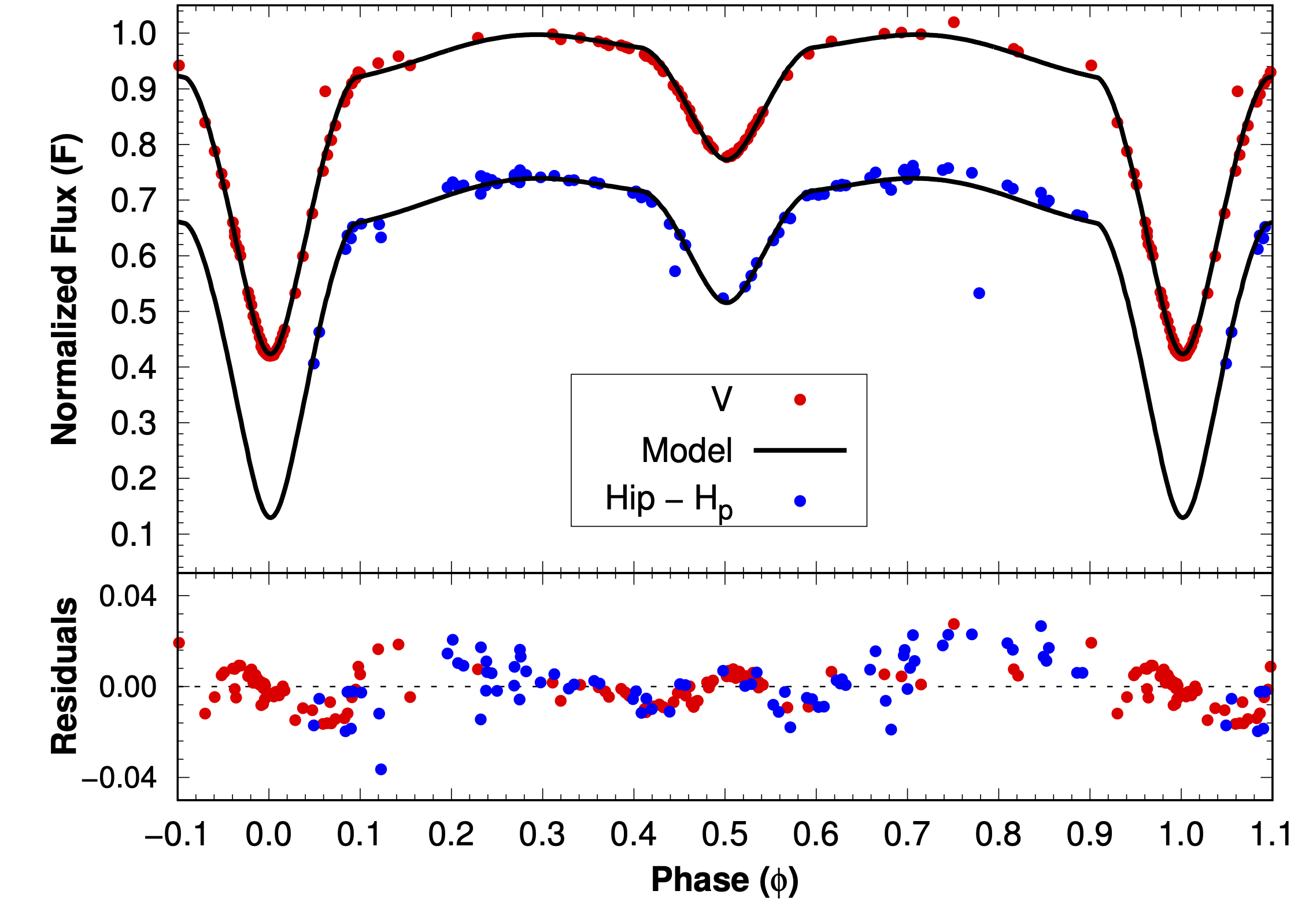}
\end{subfigure}
\newline
\begin{subfigure}[htbp]{0.8\linewidth}
\includegraphics[width=\linewidth]{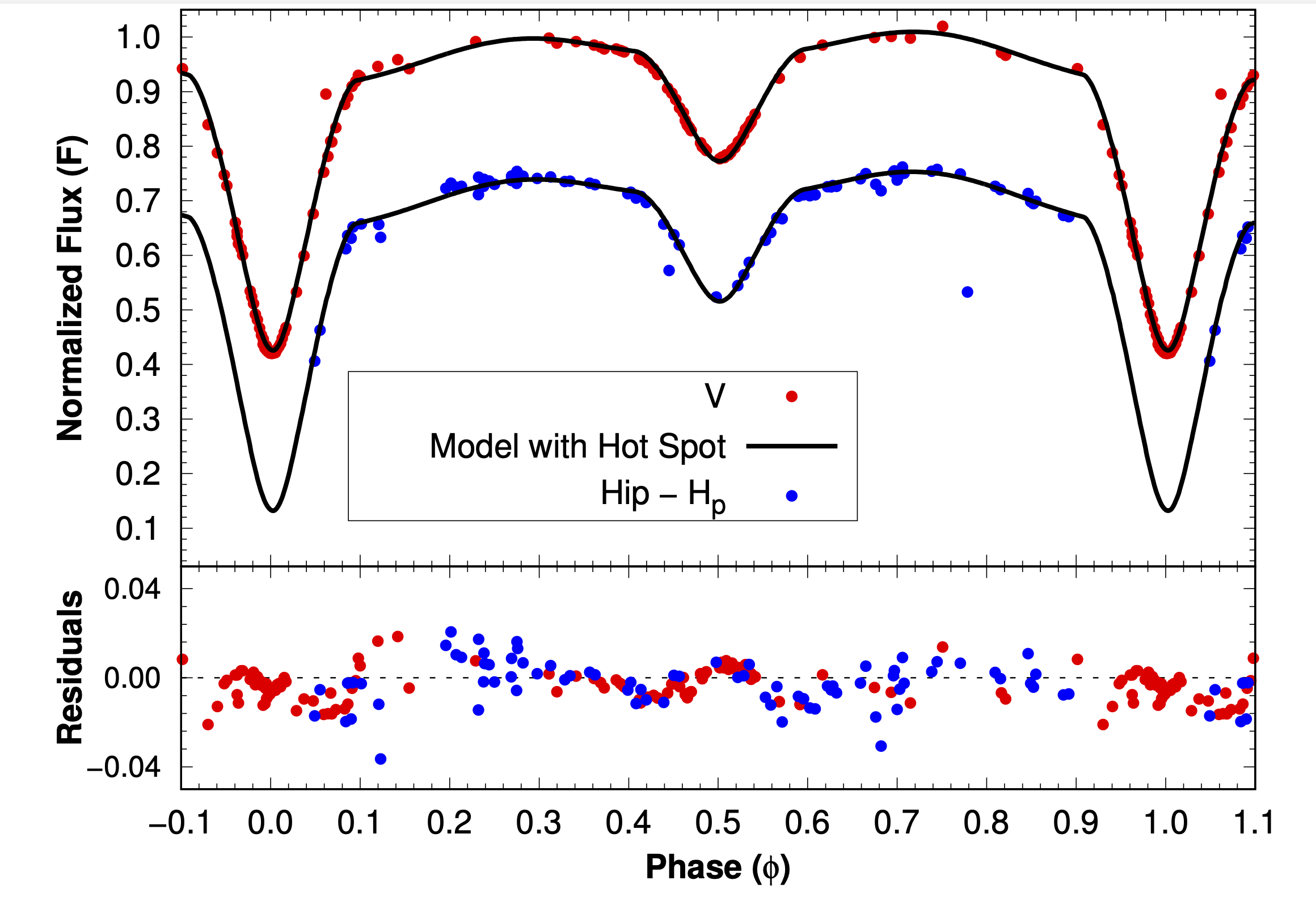}
\end{subfigure}
\newline
\begin{subfigure}[h]{0.8\linewidth}
\includegraphics[width=\linewidth]{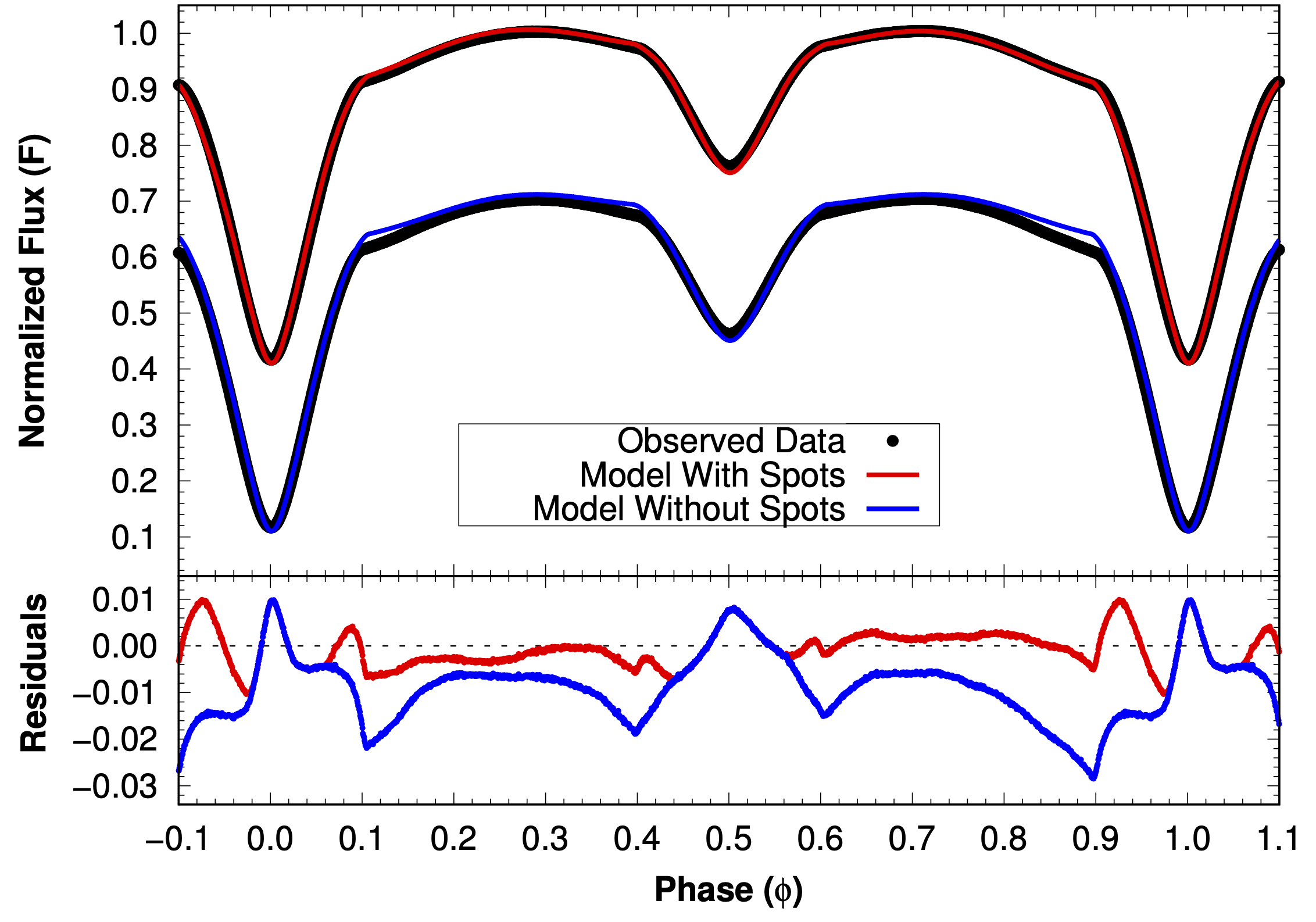}
\end{subfigure}
    \caption{Observed LCs and the models with and without hot spots of the RS~Sgr. \textit{Top panel:} The Johnson-V band and {\it HIP} LCs with models without hot spots. \textit{Middle panel:} Same as top panel but with hot spot mode. \textit{Bottom panel:} {\it TESS} LC and models with (red) and without (blue) cool spots.}
    \label{fig:LC}
\end{figure}

\begin{figure*}
    \centering
    \includegraphics[width=1\linewidth]{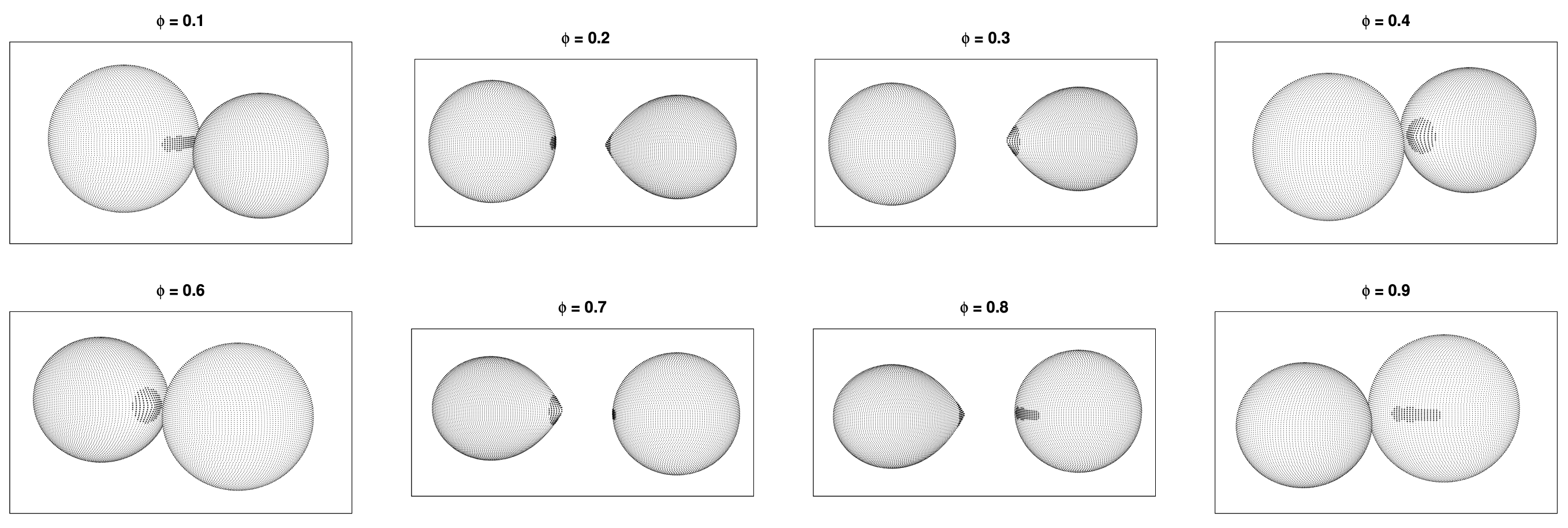}
    \caption{The Roche equipotentials for the components of RS Sgr determined from the solution of the {\it TESS} LC with the spot assumption.}
    \label{fig:roche}
\end{figure*}

\begin{figure}
   {\includegraphics[width=1\linewidth]{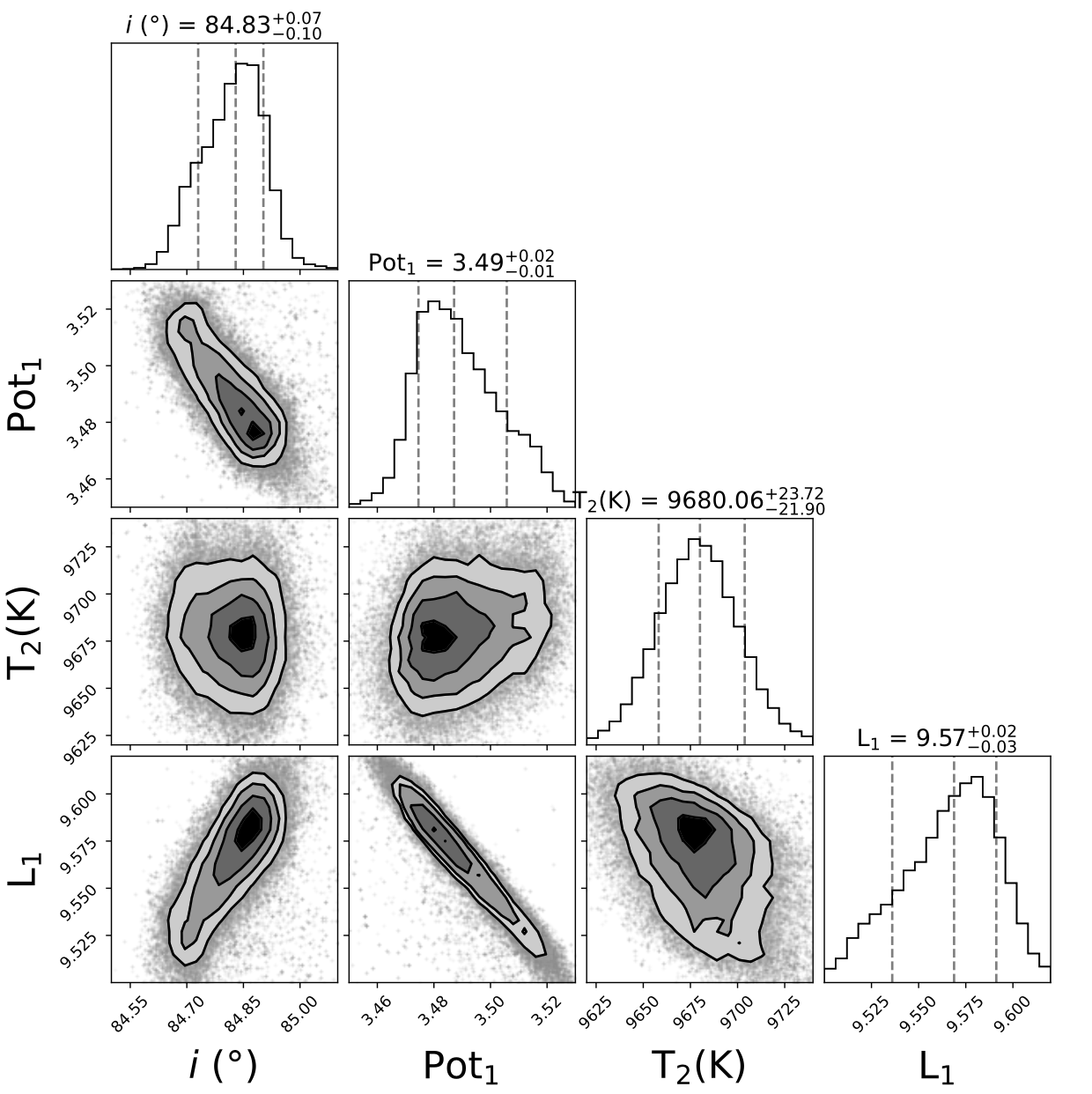}}
    \caption{The corner plots of the system parameters determined by MCMC modeling.}
    \label{fig:histogram}
\end{figure}

\begin{figure}
	\includegraphics[width=\columnwidth]{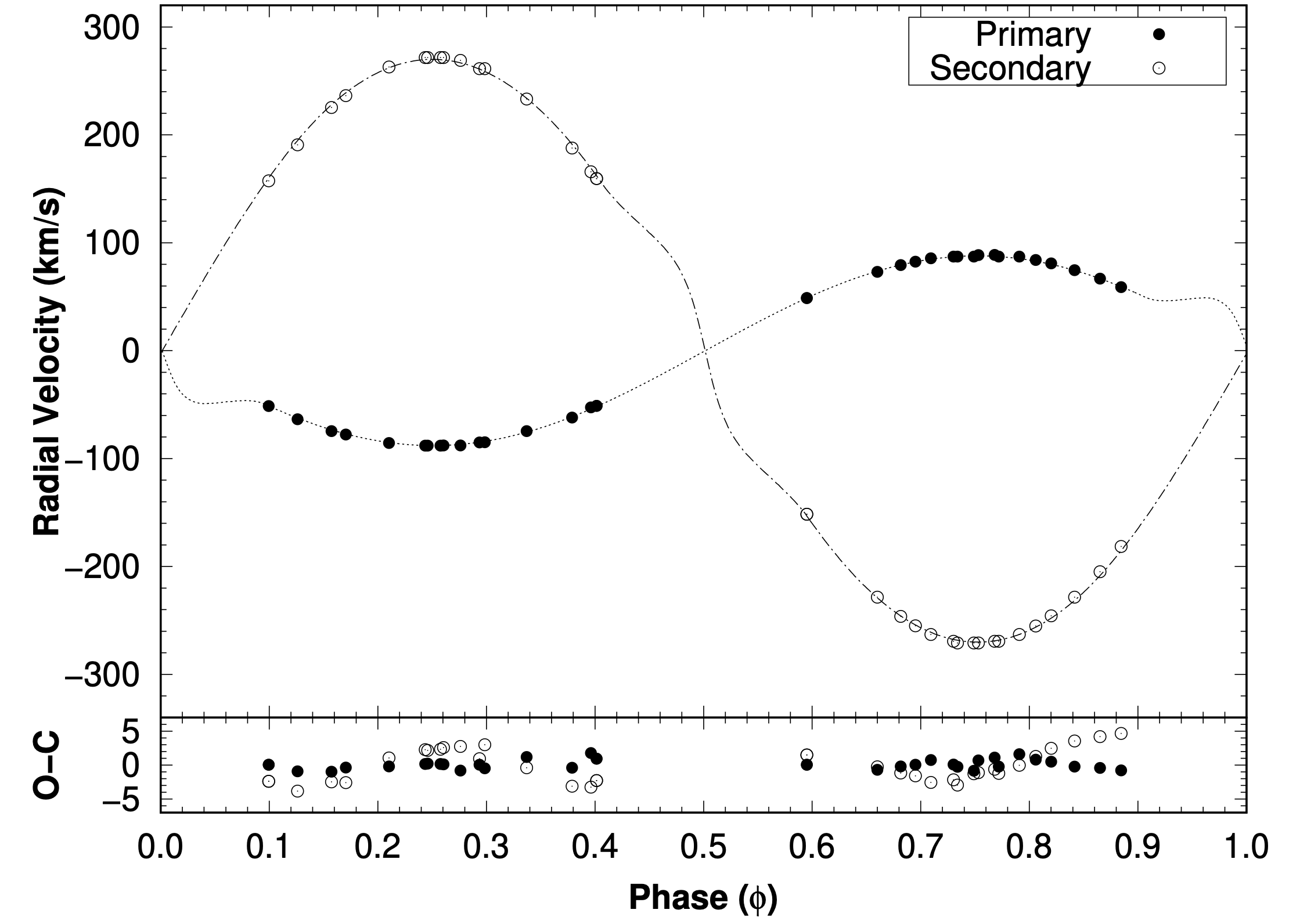}
     \caption{Observed data and the best fitting RV curves. Open and filled circles show the velocities of the secondary and the primary components, respectively. The RVs shown are relative to $V_\gamma$.}
    \label{fig:rvcurve}
\end{figure}

The observed photometric and RV data, along with the model curves based on the final parameters, are illustrated in Figures~\ref{fig:LC} and \ref{fig:rvcurve}, respectively. The system geometry is shown in Figure~\ref{fig:roche}, while Figure~\ref{fig:histogram} presents the posterior distributions of selected parameters and their representation as a multivariate Gaussian.
 
In the top and middle panels of Fig.\ref{fig:LC}, models with and without hot spots are shown for the {\it V} and {\it HIP} bands. The LC residuals of both bands exhibit a consistent pattern, indicating that the mass transfer mechanism behaves similarly and is observable in both datasets. The effect of the hot zone, formed by the impact of accreting material on the primary component, is clearly visible between orbital phases $\sim$0.75 and 0.90 in the {\it HIP} LC. A hot spot was added to the primary's surface at a longitude of 90°, with a radius of 20° and a temperature factor of 1.15. These values were adjusted to best reproduce the excess brightness observed in the {\it HIP} band between phases 0.75 and 0.90. In contrast, absorption caused by the projected flow of gas onto the components appears as brightness dips around phases 0.1, 0.4, 0.6, and 0.9 in the {\it TESS} LCs. Although the brightness decrease is not as prominent as in the {\it TESS} LC, it is still observable in the V-band data of Shobbrook (2004) around phases 0.05, 0.4, and 0.9, supporting the features seen in the {\it TESS} curves (see LC residuals panel in Fig.\ref{fig:LC}).

It is expected that the effect of the hot spot becomes weaker at longer wavelengths, such as the {\it TESS} passband, and relatively stronger at shorter wavelengths. The {\it TESS} LC was modelled by placing artificial cool spots—representing light-absorbing regions—on specific surface areas of the components. These are not physical spots but approximations used to model the observed light loss. To test whether the wavelength-dependent brightness variations could be attributed to surface reflection effects \citep[e.g. V716 Cep,][]{2008MNRAS.385..381B}, we experimented with models using different albedo values, sometimes treating albedo as a free parameter during fitting. However, no satisfactory model could be obtained, indicating that albedo variations cannot account for the observed LC features.

\begin{table}
	\caption{Results from the solution of LCs of RS~Sgr.}	
	\begin{tabular}{p{0.15\columnwidth}p{0.15\columnwidth}p{0.15\columnwidth}p{0.175\columnwidth}p{0.175\columnwidth}}
		\hline 
		\hline
                   & \multicolumn{2}{c}{ Hip and V band} &  \multicolumn{2}{c}{{\it TESS}} \\ \hline
		Parameter &  Primary & Secondary &  Primary & Secondary \\ 
         (Unit) & & & & \\
         \hline 
        
		\hline 
$P$ (d)             & \multicolumn{4}{c}{2.4156835 (fixed)}                        \\
$T_0$ (d)           & \multicolumn{4}{c}{2452502.4058 $\pm$ 0.0003 }     \\
$T_\mathrm{eff}$ (K)       &  19000 (fixed)    & 9950$\pm$ 200            &  19000 (fixed)    & 9680 $\pm$ 124                \\
$i$ ($^{\circ}$)    & \multicolumn{2}{c}{82.58$\pm$ 0.24}          & \multicolumn{2}{c}{82.829 $\pm$ 0.074}              \\
$e$                 & \multicolumn{4}{c}{0.0(fixed)}                       \\
$q$                 & \multicolumn{4}{c}{0.325 (fixed)}              \\
$L_{V}/L_T$         & 0.75 $\pm$ 0.01        & 0.25 $\pm$  0.01           &            &                 \\
$L_{Hip}/L_T$       & 0.79 $\pm$ 0.01        & 0.21 $\pm$ 0.01           &     &                 \\
$L_{Tess}/L_T$       &        &                                         & 0.762 $\pm$ 0.002 & 0.238 $\pm$ 0.002                \\
$\Omega$            &   3.76 $\pm$ 0.07 &  2.52  &  3.487 $\pm$ 0.019           & 2.52                      \\
$A_{1,2}$       & \multicolumn{4}{c}{1.0 (fixed)}\\
$g_{1,2}$       & \multicolumn{4}{c}{1.0 (fixed)}\\
$r\,(=R/a)$                 & 0.302$\pm$ 0.001      & 0.286 $\pm$ 0.001         & 0.321 $\pm$ 0.001      & 0.286 $\pm$ 0.001               \\
filling factor (f)  & \multicolumn{2}{c}{0.67} & \multicolumn{2}{c}{1.0}\\
\hline    
   &\multicolumn{4}{c}{Spot parameters of Tess LC} \\
   \hline
    &Colatitude&	Longitude&    Radius&	Temperature \\ & & & & factor  \\
    \hline 
    \multicolumn{5}{c}{Primary}  \\
    &    90&	0&	6&	0.45    \\
    &    90&	5&	5&	0.45    \\
    &    90&	10&	6&	0.45    \\
    &    90&	15&	5&	0.45    \\
    &    90&	20&	5&	0.45    \\
    &    90&	25&	5&	0.45    \\
    &    90&	30&	5&	0.45    \\
    \multicolumn{5}{c}{Secondary} \\		
    &    90&	0&	6&	0.36    \\
    &    90&	0&	13&	0.65    \\
		\hline 
		\hline
		\label{tab:LC} 
	\end{tabular}
\end{table}

 
\section{Stellar Atmosphere Models}

Following the photometric and spectroscopic analyses, including LC modelling and atmospheric fitting, further refinement of the stellar parameters was pursued through the spectral disentangling approach. Although the temperatures and surface gravities of the components in binary systems can be estimated through photometric analysis, spectroscopic determination of these parameters generally yields more precise and reliable results. For this reason, the disentangled spectra of the components—obtained using the \textsc{\large korel} code—were employed in the present study. This method has been successfully applied to similar systems such as $\delta$ Lib \citep{2006MNRAS.370.1935B}, MQ Cen \citep{2016cosp...41E1950T,MQcenBakis}, and V716~Cep \citep{2008MNRAS.385..381B}.

To determine the atmospheric parameters from the disentangled spectra, an iterative fitting procedure was applied using the initial values of $T_\mathrm{eff}$, $\log g$, and $v_\mathrm{rot}$ derived from the preliminary photometric solution. During this process, synthetic spectra were computed for various parameter sets and compared to the observed disentangled spectra. The final atmospheric parameters were adopted based on the best-fitting model, corresponding to the combination of effective temperature, surface gravity, and rotational velocity that minimised the residuals and yielded the most consistent spectral match for each component.

For the spectral modelling, we employed the \textsc{\large tlusty200} code \citep{tlustyhubeny,Hubeny1995}, which generates model atmospheres under non-LTE conditions for stars with $T_\mathrm{eff} > 10000$K, together with the \textsc{\large synspec49} code to compute the corresponding synthetic spectra. As previously inferred from the system's photometric colours, the presence of relatively strong neutral helium lines in the optical spectrum of RS~Sgr further supports the classification of the primary component as an early-type star.

Model atmosphere grids and synthetic spectra were constructed assuming solar abundances, covering an effective temperature range of 17000–22000K (in steps of 100K) and surface gravity ($\log g$) values from 3.7 to 4.3 (in steps of 0.1). For each grid point, synthetic spectra were calculated and compared to the disentangled spectra of the primary component using a chi-square minimisation method. This statistical comparison identified the parameter set that best matched the observed spectra across all three spectral orders analysed (97, 113, and 127).

\begin{figure*}
    \centering
    \includegraphics[width=1\linewidth]{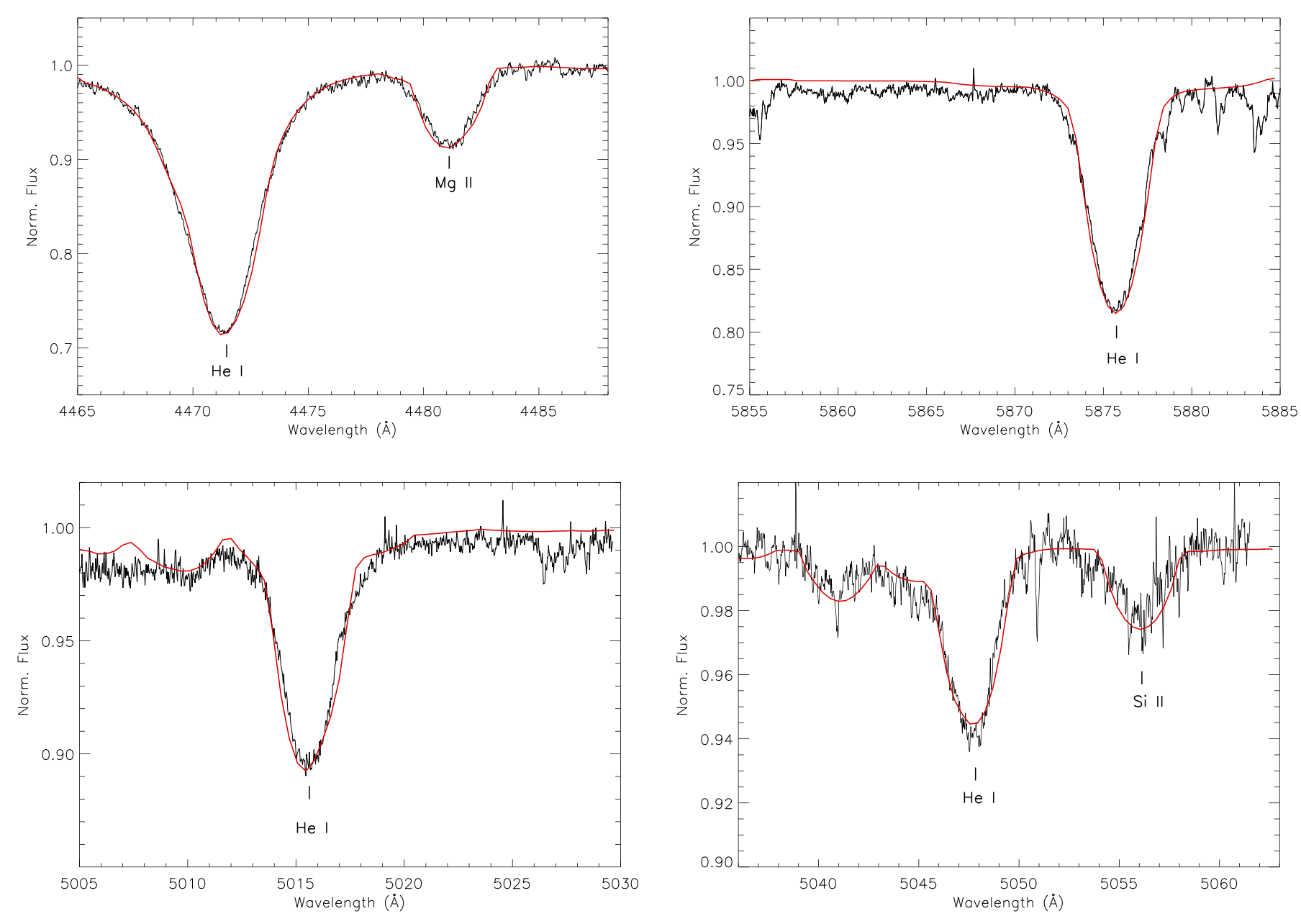}
    \caption{The best-fitting synthetic model spectra of the primary component. Black and red lines show the observed and the synthetic model spectra generated by the \textsc{\large tlusty200} code, respectively.}
    \label{primarymodel}
\end{figure*}

The best-fitting atmospheric parameters for the primary component were determined to be $T_\mathrm{eff,1} = 19000$~K, $\log g_1 = 4.0$, and $v \sin i_1 = 115\,$kms$^{-1}$. The disentangled spectra of the primary component for orders 97, 113, and 127, along with the corresponding best-fitting synthetic model spectra, are presented in Figure\ref{primarymodel}.

Since the temperature of the secondary component, as derived from photometric modelling, lies below 10000~K, the model atmosphere code \textsc{\large atlas9}—which operates under the assumption of local thermodynamic equilibrium (LTE)—was utilised for this star. The corresponding synthetic spectra were computed using the \textsc{\large synthe} code \citep{castelli1988kurucz, Kurucz1970}.

Model atmosphere grids assuming solar composition were prepared for effective temperatures ranging from 9000 to 10000 K (in 100 K steps) and surface gravities ($\log g$) between 3.0 and 3.5 (in 0.1 dex steps). For each parameter set, synthetic spectra were calculated and compared with the disentangled spectrum of the secondary component, using the same chi-square minimisation technique applied for the primary.

\begin{figure*}
    \centering
    \includegraphics[width=1\linewidth]{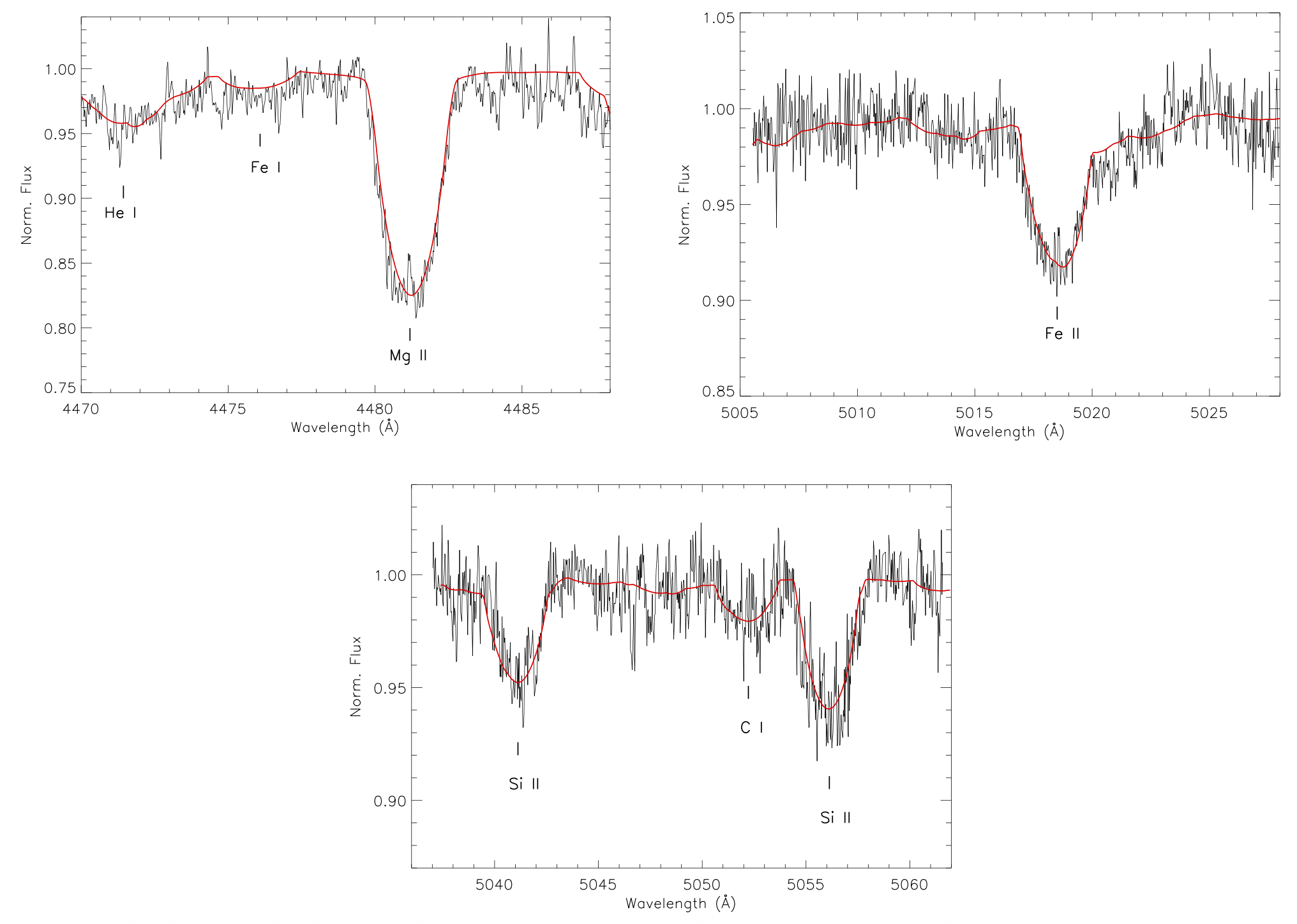}
    \caption{The best fitting synthetic model spectra of the secondary component. Black and red lines show observed and the synthetic model spectra generated by the \textsc{\large atlas9} code, respectively.}\label{secondarymodel}
\end{figure*}

The best-fitting model parameters for the secondary component were found to be $T_\mathrm{eff,2} = 9500$ K, $\log g_2 = 3.5$, and $v \sin i_2 = 90\,$kms$^{-1}$. For both the LTE and non-LTE models, microturbulence velocity and the instrumental broadening full-width at half maximum (FWHM) were adopted as 2 kms$^{-1}$ and 0.025\,{\AA}, respectively. The disentangled spectrum of the secondary component, along with the best-fitting synthetic spectrum, is displayed in Figure~\ref{secondarymodel}. A summary of the final model parameters for both components of RS~Sgr is presented in Table\ref{tab:rssgrmodel}.

\begin{table}
	\centering
	\caption{\textsc{\large tlusty200} and \textsc{\large atlas9} model atmosphere parameters of  RS~Sgr.}
	\begin{tabular}{lccc}
		\hline 
		\hline
		Parameters & Symbol & Primary & Secondary \\ 
		\hline 
		\hline 
		Effective Temperature (K)& $T_{\mathrm{eff}}$  & 19000 $\pm$ 100 & 9500 $\pm$ 100  \\
		Surface Gravity (cgs) & $\log{g}$ & 4.0 $\pm$ 0.1 & 3.5 $\pm$ 0.1 \\
		Rotational Velocity (km s$^{-1}$) & $vsini$ & 115 $\pm$ 5 & 90 $\pm$ 5\\
		Metallicity & $[M/H]$& 0.0 & 0.0 \\
		\hline 
		\hline 
		\label{tab:rssgrmodel}
	\end{tabular}

\end{table}

\section{Astrophysical parameters and the distance of the system}

Following the detailed determination of the atmospheric parameters for both components, the next step was to evaluate the absolute parameters of the system based on the simultaneous modelling of the light and RV curves.
The absolute parameters from the simultaneous solution of the light and RV curves are summarised in Table \ref{tab:absoluteparam}. According to Table \ref{tab:absoluteparam}, the separation between the components is 17.21 $R_{\odot}$ and the masses of the components for the primary and the secondary are 8.85 $M_{\odot}$ and 2.88 $M_{\odot}$, respectively. Using calibration tables of \cite{straivzys1981fundamental}, the mass, the radius ($R_1$ = 5.56 $R{\odot}$), and the temperature (19000 K) of the primary component refer to the B2.5 V, B1.5 V, and B3V spectral type, respectively. The parameters of the primary component give a spectral type that is approximately compatible with each other. The mass, radius ($R_2$ = 4.93 $R{\odot}$), and temperature (9700 K) of the secondary component indicate B8.5 IV / B9 III, B3 IV / B4 III, and A0 III spectral types, respectively. The temperatures obtained from the solution of the light and RV curves seem to be compatible with those determined from the atmosphere models. Therefore, based on the derived astrophysical parameters from atmosphere modelling and simultaneous LC and RV solutions, RS~Sgr is composed of two early-type stars with B3V+A0III spectral type. In semi-detached systems, the primary component, the mass gainer, has a lower luminosity, while the component that loses mass (donor) has a greater luminosity compared to their masses \citep{ibanoglu2006angular}. This situation actually emerges when it is considered the evolution of the components under the assumption of single-star evolution.

In addition to the dynamical and atmospheric analysis, the distance estimation offers an independent verification of the derived parameters. The GAIA \citep{Gaia2016} DR3 \citep{GaiaDR3_2023} parallax measurement ($\varpi=$2.2918$\pm$0.0977 mas) locates RS~Sgr at a distance of $d= 436^{+19}_{-18}$ pc. If we adopt $E(B-V)=$0.122 mag from the $Q$-method (see \S4), the interstellar extinction in the $V$-band becomes $A_V$ = 3.1$\times$E(B-V) = 0.378 mag. The absolute visual magnitudes of the components were calculated from the bolometric magnitude and the bolometric correction (BC) values given by \cite{Eker2020}. Using the absolute magnitude of the components with the $V$-magnitude of the system (Table \ref{tab:absoluteparam}), light contributions of the components (Table \ref{tab:LC}) and the interstellar extinction value, the distance to the system comes out to be 418 $\pm$ 15 pc, which is $\sim$18 pc away from the GAIA determination, but still within the 1$\sigma$ error range.

\begin{table*}
	\centering
	\caption{Absolute parameters of the  RS~Sgr. Errors of calculated parameters are given in parentheses.}
	\label{tab:absoluteparam}
	\begin{tabular}{llcccc}
		\hline 
		\hline
  & & \multicolumn{2}{c}{Hip and V filter} & \multicolumn{2}{c}{TESS}\\
  \hline
		Parameter & Symbol & Primary & Secondary & Primary& Secondary \\
		\hline 
		\hline 
Spectral type			&	Sp			&	B3 V		&	A0 III		&	B3 V		&	A0 III		\\
Mass (M$_\odot$ )		&	\textit M		&	8.85 $\pm$ 0.03  &	2.88 $\pm$ 0.02	&	8.846 $\pm$ 0.015	&	2.875 $\pm$ 0.012 \\
Radius (R$_\odot$)		&	\textit R		&	5.23 $\pm$ 0.02	&	4.95 $\pm$ 0.02	&	5.524 $\pm$ 0.022	&	4.929 $\pm$ 0.014	\\
Separation (R$_\odot$)		&	\textit a		&	\multicolumn{4}{c}{17.32 $\pm$ 0.02}		\\
Surface gravity (cgs)		&	\textit logg		& 3.97 $\pm$ 0.01	&	3.52 $\pm$ 0.01	&	3.900 $\pm$ 0.005	&	3.450 $\pm$ 0.003	\\
Integrated visual magnitude (mag)	&	\textit V	\textit Tess	&	\multicolumn{2}{c}{6.03 $\pm$ 0.03} & \multicolumn{2}{c}{6.24 $\pm$ 0.01}		\\
Individual visual magnitudes (mag)	& 	\textit V$_{1,2}$ \textit Tess$_{1,2}$	&	6.34 $\pm$ 0.03	&	7.54 $\pm$ 0.03	& 6.54 $\pm$ 0.01 &	7.80 $\pm$ 0.01	\\
Integrated color index (mag)	&	\textit B-V		&	\multicolumn{4}{c}{-0.08}			\\
Temperature (K)			&	\textit T$_\mathrm{eff}$	&	19000 $\pm$ 100 &	9950 $\pm$ 200	&	19000 $\pm$ 100	&	9680 $\pm$ 124	\\
Luminosity (L$_\odot$)		&	\textit logL		&	3.51 $\pm$ 0.01	&	2.34 $\pm$ 0.04	&	3.57 $\pm$ 0.01	&	2.29 $\pm$ 0.03	\\
Bolometric magnitude (mag)	&	\textit M$_{bol}$	&	-4.02 $\pm$ 0.03	&	-1.09 $\pm$ 0.10	&	-4.16 $\pm$ 0.03	&	-0.97 $\pm$ 0.07\\
Absolute visual magnitude (mag)	&	\textit M$_{v 1,2}$	\textit M$_{tess 1,2}$	&	-2.23 $\pm$ 0.04	&	-0.85 $\pm$ 0.10	&	2.28 $\pm$ 0.04	&	-0.70 $\pm$ 0.07\\
Bolometric correction (mag)	&	\textit BC		&	-1.79 $\pm$ 0.01	&	-0.24 $\pm$ 0.04	& -1.88 $\pm$ 0.01&	-0.27 $\pm$ 0.01	\\
Computed synchronization  velocities (km s$^{-1}$)	&	\textit v$_{synch}$	&	110 $\pm$ 1		&	104 $\pm$ 1		&	117 $\pm$ 1		&	104 $\pm$ 1		\\
vsini (km s$^{-1}$)	&	\textit v$_{rot 1,2}$	& 115 $\pm$ 5 & 90 $\pm$ 5 & 115 $\pm$ 5 & 90 $\pm$ 5	\\

		\hline 
		\hline 
			\end{tabular}
\end{table*}
	
\section{Mass Transfer and Accretion}

\begin{figure}
	\includegraphics[width=\columnwidth]{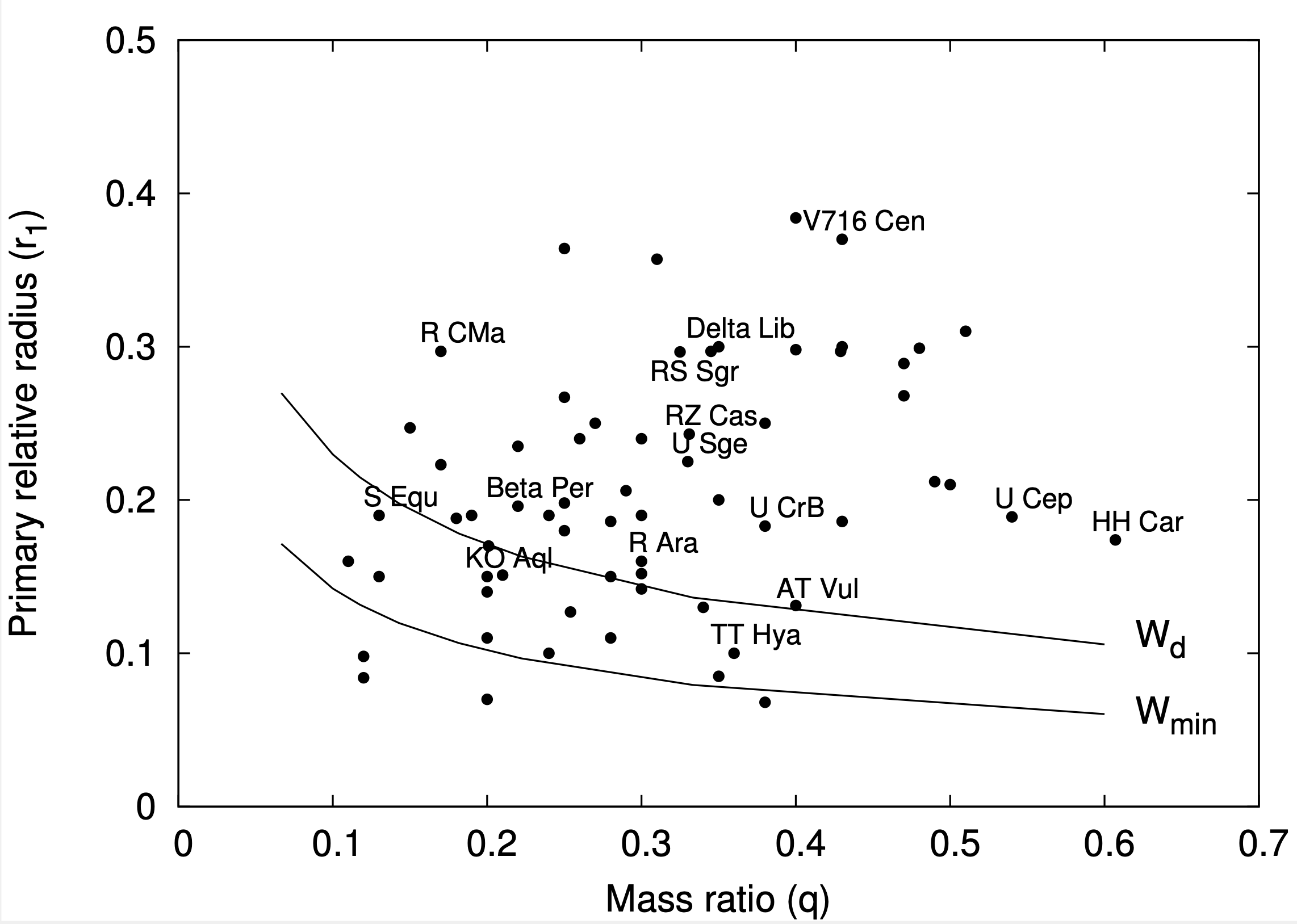}
    \caption{The positions of the  RS~Sgr and some Algols in the r$_1-$q  diagram.}
    \label{fig:r1q}
\end{figure}

Theoretical models enable the prediction of how material transferred from the mass-losing component in an interacting binary system evolves, depending on the fractional radii of the components, their mass ratio, the escape velocity, and the direction of the flow. These models can describe whether the transferred matter will directly accrete onto the mass-gaining star or form a transient or stable accretion structure such as a disk. \cite{lubow1975gas} conducted a foundational theoretical study on the trajectories of gas flow and the structures that can form around the mass-gaining star in close binary systems undergoing mass transfer. Later, \cite{1989SSRv...50....9P} introduced the (r$1$–q) diagram for interacting binaries with known fractional radii of the primary component ($r_1$) and mass ratios ($q$). In Figure \ref{fig:r1q}, W${\rm d}$ represents the boundary for the formation of a classical accretion disk for different mass ratios, while W$_{\rm min}$ denotes the minimum approach distance of the gas stream to the surface of the accretor.

According to \cite{lubow1975gas}, this diagram consists of three regions: Region I lies below the W$_{\rm min}$ curve. In systems located in this region, the gas stream does not impact the mass-gaining component directly; instead, it forms a stable accretion disk that is continuously fed by the transferred material. Region II lies between the W${\rm min}$ and W${\rm d}$ curves. Systems in this region can simultaneously possess a permanent accretion disk and a shock zone where the gas stream collides with the surface of the mass gainer. Region III lies above the W$_{\rm d}$ curve. In this region, the gas stream impacts the mass-gaining component directly, resulting in the formation of an impact (shock) region. A transient accretion disk may also form around the primary component as a secondary effect.

According to the position of RS~Sgr in the $r_1$–$q$ diagram, with a fractional radius of the primary component ($r_1 = R_1/a = 0.295$) and a mass ratio of $q = 0.311$, the system is located well above the transient disk boundary. Therefore, it is expected that the gas stream directly strikes the surface of the primary star, leading to the formation of a prominent shock zone.

The gas stream and the resulting impact zone are expected to influence both the photometric and spectroscopic data of the system. To investigate this, we first examined the difference between the observed and synthetic LCs of the system. For this analysis, {\it TESS} data were utilised due to their higher precision and superior phase coverage.

\begin{figure}
	\includegraphics[width=\columnwidth]{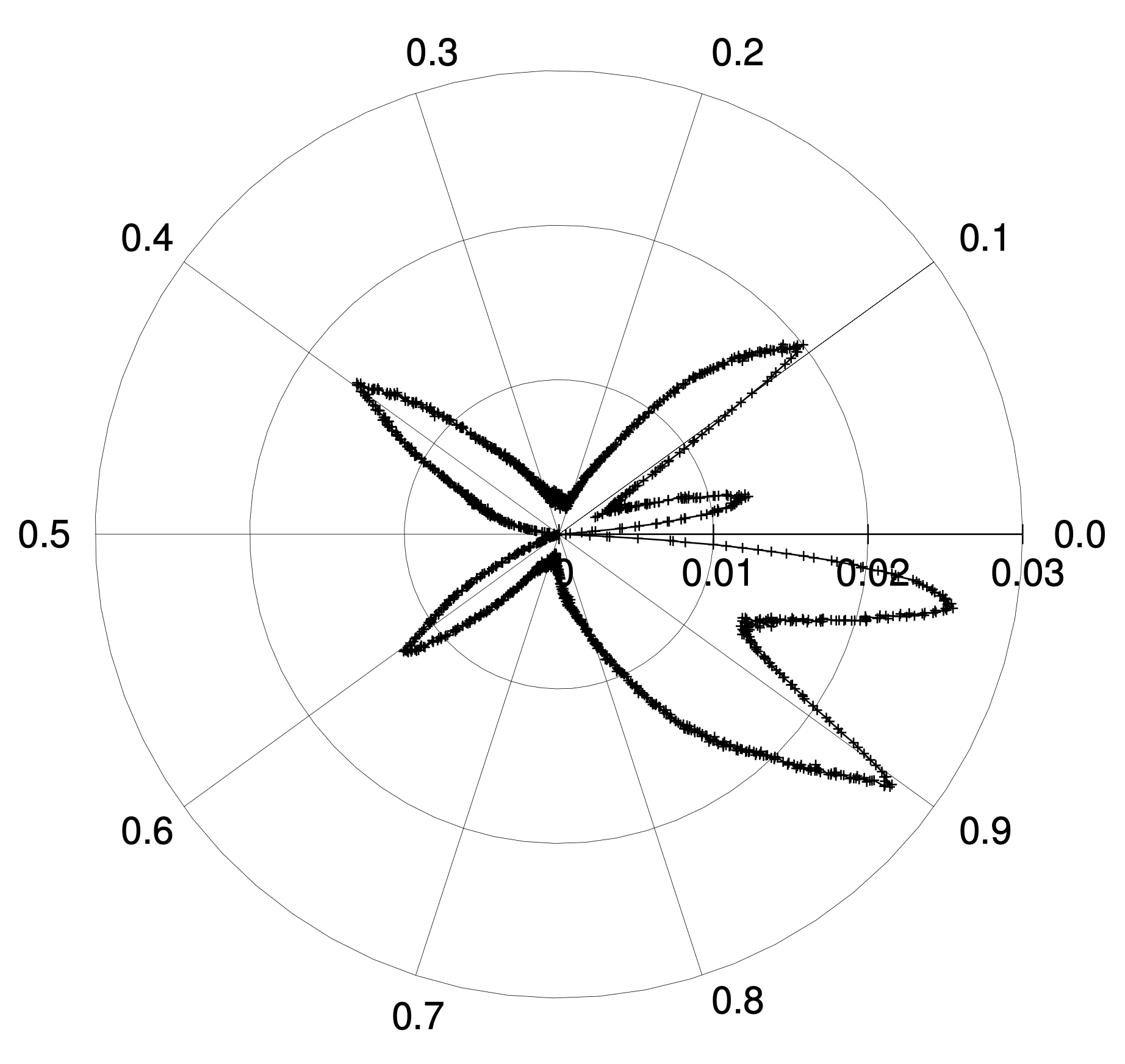}
    \caption{It is a phase-residual graph created by {\it TESS} photometric data and the synthetic LC without spots of the  RS~Sgr system. The variation of the optical depth of the material (radial axis) with the orbital phase (numbers around the circle) is shown.}
    \label{fig:3D}
\end{figure}

The optical depth of the circumstellar material can be estimated using the relation $\tau$ = $\ln(I_{0}/I)$, where $I$ is the observed normalised flux and $I_{0}$ is the synthetic (model) normalised flux. Figure \ref{fig:3D} shows the distribution of the circumstellar material based on this calculation. The figure indicates that the circumstellar matter is most prominent in the phase interval 0.8 -- 0.0, while it appears to be less dominant between phases 0.1 and 0.4.

Similar to the photometric data, the spectroscopic data should also exhibit signatures of the impact region and the gas stream within the system. In short-period Algol-type systems like RS~Sgr, the emission and absorption features caused by accretion around the hotter component typically contribute only weakly to the total system spectrum, being dominated by the two stellar components.

\begin{figure}
	\includegraphics[width=0.8\columnwidth]{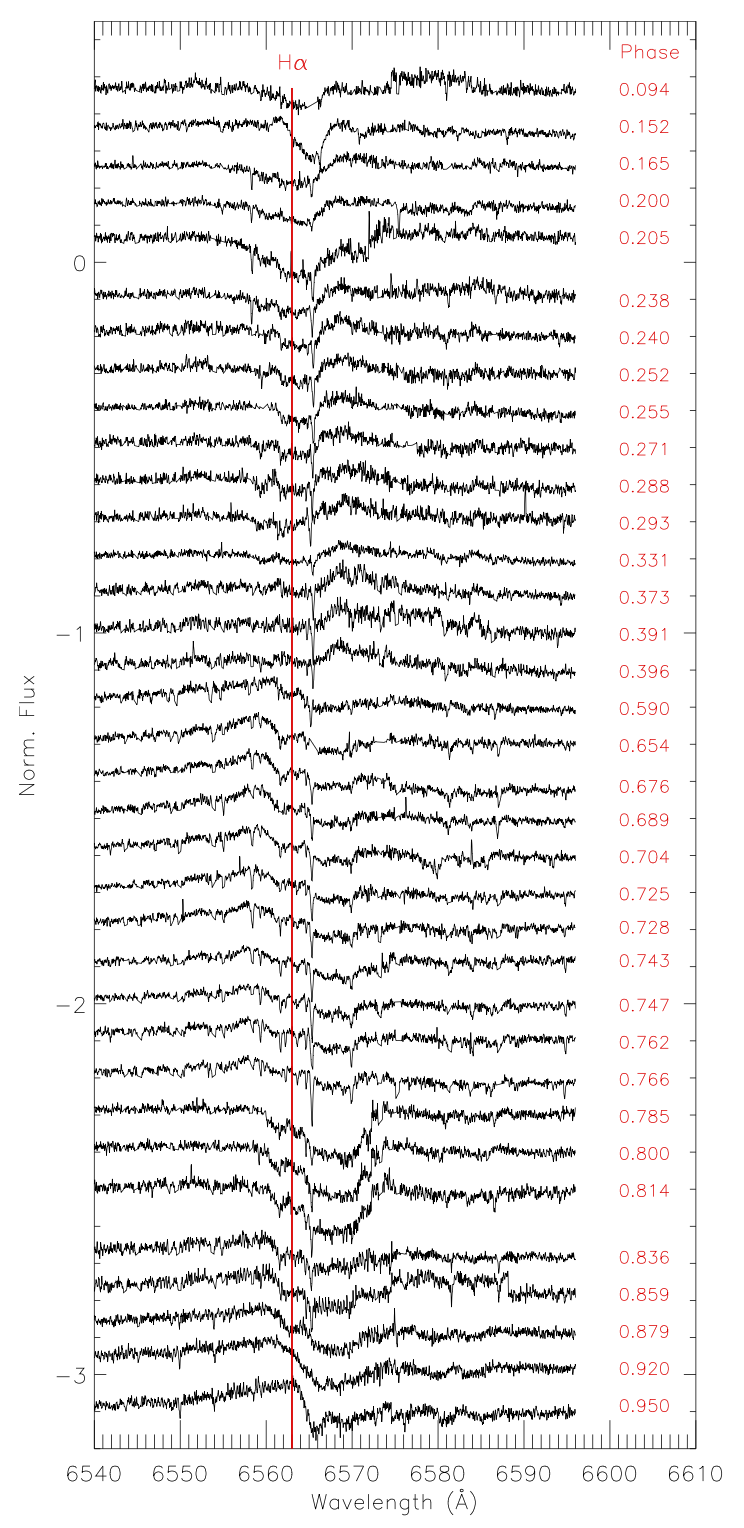}
    \caption{The difference spectra show the extra absorption and emission characteristics. The orbital phases are shown on the right. The vertical red line represents the laboratory wavelength of the H$_\alpha$ line ($\lambda_0$=6562.82 {\AA})}
    \label{fig:difference}
\end{figure}

To isolate the contribution from other physical processes—such as magnetic activity, accretion flows, or the gas stream—the light contribution of the stellar components in the $H_\alpha$ region is subtracted from the observed spectra. The resulting residuals are referred to as difference spectra. In constructing these spectra, the surface temperature of each component is assumed to be uniform, following the approach of \citet{mercedes1999}, although in reality, a minor temperature gradient exists due to the Roche geometry of the system. Figure \ref{fig:difference} presents the difference spectra sorted by orbital phase.

\begin{figure}
	\includegraphics[width=1\columnwidth]{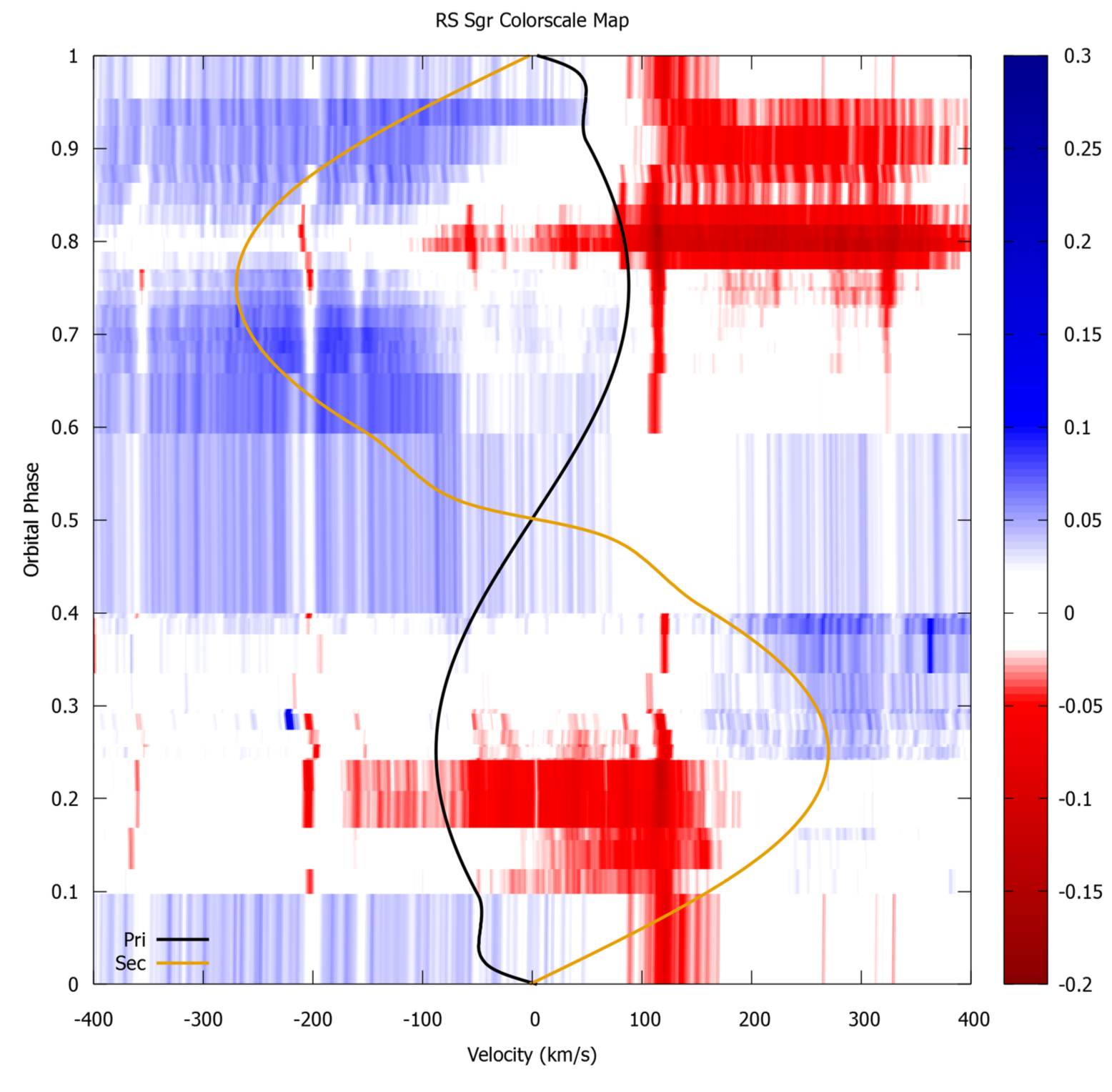}
    \caption{Colorscale image created by the difference H$_\alpha$ line profiles of  RS~Sgr. Red areas indicate increased residual absorption.}
    \label{fig:colorscale}
\end{figure}

Using the difference profiles, a trailing spectrogram illustrating the orbital motion of the components has been constructed, as shown in Fig.~\ref{fig:colorscale}. This velocity-space diagram highlights the regions of emission and absorption throughout the orbital cycle.

From Fig.~\ref{fig:colorscale}, weak emission is observed between orbital phases $\varphi=0.3$–$0.4$ and $\varphi=0.6$–$0.75$, spanning a wide range of velocities. This feature likely originates from an additional emission or absorption region caused by the impact of the gas stream onto the primary component. Such an interaction may lead to the temporary formation of an accretion structure between the primary and secondary stars.

Additionally, an absorption feature is visible between $\varphi=0.15$–$0.25$ and $\varphi=0.80$–$0.85$, redshifted with respect to the systemic velocity. This likely corresponds to material flowing from the secondary component through the first Lagrangian point (L$_1$), moving away from the observer.

Another notable absorption feature, which appears to co-move with the primary star between phases $\varphi=0.75$–$0.85$, may be attributed to the stream of accreting gas. This interpretation is supported by the structures seen in Fig.\ref{fig:3D}, Fig.\ref{fig:difference}, and Fig.~\ref{fig:colorscale}.

\section{Conclusions}

In this study, archival high-resolution spectra and available LCs from the literature were analyzed for RS~Sgr, a classical Algol-type interacting binary system. The primary aim was to derive precise astrophysical parameters of the stellar components and to investigate the mass transfer and accretion structures around the hotter, mass-gaining component through H$_\alpha$ difference profiles.

For the first time, the fundamental spectroscopic and absolute parameters of RS~Sgr have been determined with high precision. Using spectral disentangling and synthetic model fitting, the physical properties of the components—effective temperatures, surface gravities, RV semi-amplitudes, and projected rotational velocities—were derived. The primary component was identified as a B3 main-sequence star with $T_{\rm eff} = 19000$ K, while the secondary component was found to be an A0-type star with $T_{\rm eff} = 9700$ K, having recently evolved off the main sequence.

The secondary component, which fills its Roche lobe, is transferring mass through the inner Lagrangian point (L$_1$) to the hotter primary component. The projected rotational velocities of both stars are consistent with synchronous rotation, indicating that the impact of the accreting gas does not significantly spin up the primary—consistent with expectations for Algol systems with short orbital periods.

Analysis of the H$_\alpha$ difference spectra revealed complex circumstellar structures. Both emission and absorption features support the presence of a low-density accretion disk and a localized impact region (hot spot) where the transferred material interacts with the primary. These features are phase-dependent and align with structures identified in the system’s photometric residuals.

The distance to RS~Sgr was estimated to be $418 \pm 15$ pc, which is slightly offset from the value derived from GAIA DR3 parallax measurements, but still within a reasonable margin of error. Importantly, the parameters derived independently from photometric and spectroscopic analyses are in good agreement, reinforcing the robustness of the results.

Overall, RS~Sgr serves as a compelling example of mass-transfer dynamics and accretion phenomena in classical Algol systems, and this study contributes significantly to the understanding of its current evolutionary stage and circumstellar environment.

\begin{acknowledgement}

We thank the anonymous referee for their insightful and constructive suggestions, which significantly improved the paper. We thank Prof. John Hearnshaw for providing observation time with HERCULES and Prof. Edwin Budding for his help during the observations. This work uses the VizieR catalog access tool, CDS, Strasbourg, France; the SIMBAD database, operated at CDS, Strasbourg, France. This work presents results from the European Space Agency (ESA) space mission, Gaia. Gaia data are being processed by the Gaia Data Processing and Analysis Consortium (DPAC). Funding for the DPAC is provided by national institutions, in particular, the institutions participating in the Gaia MultiLateral Agreement (MLA). The Gaia mission website is https://www.cosmos.esa.int/gaia. The Gaia archive website is https://archives.esac.esa.int/gaia. This paper includes data collected with the TESS mission, obtained from the MAST data archive at the Space Telescope Science Institute (STScI). Funding for the TESS mission is provided by the NASA Explorer Program. STScI is operated by the Association of Universities for Research in Astronomy, Inc., under NASA contract NAS 5–26555. 
\end{acknowledgement}

\paragraph{Funding Statement}

This study is supported by The Scientific $\&$ Technological Research Council of Turkey (TUBITAK) under project code 2214. 
\paragraph{Competing Interests}

None.

\paragraph{Data Availability Statement}

The data that support the findings of this study are available from the corresponding author, HB, upon reasonable request.

\printendnotes

\printbibliography

\appendix

\end{document}